# Causal Wave Mechanics and the Advent of Complexity.
# II. Dynamic uncertainty in quantum systems and the correspondence principle

A.P. KIRILYUK[*]

Institute of Metal Physics, Kiev, Ukraine 252142

ABSTRACT. The intrinsic multivaluedness of interaction process, revealed in Part I of this series of papers, is interpreted as the origin of the true dynamical (in particular, quantum) chaos. The latter is causally deduced as unceasing series of transitions, dynamically probabilistic by their origin, between the equally real, but incompatible 'realisations' (modes of interaction) of a system. The obtained set of realisations form the causally derived, intrinsically complete "space of events" providing the crucial extension of the notion of probability and the method of its first-principle calculation. The fundamental dynamic uncertainty thus revealed is specified for Hamiltonian quantum systems and applied to quantum chaos description in periodically perturbed systems. The ordinary semiclassical transition in our quantum-mechanical results leads to exact reproduction of the main features of chaotic behaviour of the system known from classical mechanics, which permits one to 're-establish' the correspondence principle for chaotic systems (inevitably lost in any their conventional, single-valued description). The causal dynamical randomness in the extended quantum mechanics is not restricted, however, to semiclassical conditions and generically occurs also in essentially quantum regimes, even though partial "quantum suppression of chaos" does exist and is specified in our description, as well as other particular types of the quantum (truly) chaotic behaviour.

NOTE ON NUMERATION OF ITEMS. We use the unified system of consecutive numbers for formulas, sections, and figures (but *not* for literature references) throughout the full work, Parts I-V. If a reference to an item is made outside its "home" part of the work, the Roman number of this home part is added to the consecutive number: 'eq. (12)' and 'eq. (12.I)' refer to the same, uniquely defined equation, but in the second case we know in addition that it can be found in Part I of the work.

---

[*]Address for correspondence: Post Box 115, Kiev - 30, Ukraine 252030.
 E-mail address: kiril@metfiz.freenet.kiev.ua

## 2.3. Quantum chaos as dynamic multivaluedness

In part I of this work it was shown that, being reformulated in the form of the modified Schrödinger equation with the unreduced effective potential (EP), a problem is naturally splitted into multiple realisations, at least for some ranges of parameter values. Now we are going to demonstrate (see also the next section) that this splitting results in what can be interpreted as dynamical chaos:

it has the same qualitative manifestations including, especially, the intrinsic dynamical randomness for the systems with chaotic classical counterparts;

it reproduces, within the quantum-mechanical description, the classically well-established transition chaos-regularity and its point in the parameter space, as well as the phenomenon of asymptotically weak chaos in the domain of global regularity;

it provides the normal semiclassical transition for chaotic systems in full agreement with the conventional correspondence principle;

it naturally reveals the fractal structure of quantum chaos (section 4.III), also in agreement with the existing general and particular results;

and finally, it gives also several forms of the "quantum suppression of chaos" which is not absolute, however, depends on the parameters and is compatible with the existence of the 'true' dynamical chaos also in essentially quantum systems far from the semiclassical transition.

So we suppose that we are in a range of parameters giving multiple realisations $\Re_i$ ($i = 1, 2, ..., N_\Re > 1$), each of them being characterised, according to the definition, eq. (12.I), by its own complete set of eigenvalues $\{\varepsilon_{\sigma n}\}$, eigenfunctions $\{\psi_{0n}(\mathbf{r}_\sigma)\}$, branch of the effective potential $V_{\text{eff}}(\mathbf{r}_\sigma)$ and that of a measurable function, e. g. the PDD (probability density distribution), $\rho(\mathbf{r})$. Since all the realisations have a priori 'equal rights', and, from the other hand, one can always observe only one of them (because of their completeness), then the full experimentally measured PDD, $\rho_{\text{ex}}(\mathbf{r})$, should be a result of the *probabilistic superposition* of the PDD's, $\rho_i(\mathbf{r})$, for the independent individual realisations:

$$\rho_{\text{ex}}(\mathbf{r}) = \sum_{i=1}^{N_\Re} {}^\oplus \rho_i(\mathbf{r}) , \qquad (16a)$$

where $\sum^\oplus$ is the probabilistic sum of independent functions (random processes).[*)] Correspondingly, the expectation value of $\rho_{\text{ex}}(\mathbf{r})$, averaged over a large enough ensemble of identical repetitions of the same experiment, is obtained as (we use the same notation for it)

---

[*)] Note that this natural *emergence* of probability in our approach provides a fundamental extension of the corresponding notion itself (see section 6.III for a more detailed discussion).

$$\rho_{\text{ex}}(\mathbf{r}) = \sum_{i=1}^{N_\Re} \alpha_i \rho_i(\mathbf{r}) , \qquad \sum_{i=1}^{N_\Re} \alpha_i = 1 , \tag{16b}$$

where the $i$-th realisation probability, $\alpha_i$, does not, a priori, depend on $i$: $\alpha_i = 1/N_\Re$. However, as has been shown above, in many real situations (mostly within the semiclassical case) the realisations may be closely separated forming a kind of quasi-continuous distribution with varying density such that individual realisations cannot be discerned experimentally. It is thus reasonable, in a general case, to designate by index $i$ the number of a discernible realisation group with a suitable physical size, and then $\alpha_i$ can show a pronounced dependence on $i$:

$$\alpha_i = N_i/N_\Re,$$

$N_i$ being the number of realisations within the $i$-th group. It leads, eventually, to the reduction of the sum in eq. (16b) to integral with the help of the *density of realisations* analogous to the well-known density of states:

$$\rho_{\text{ex}}(\mathbf{r}) = \int_{\Omega_\iota} \delta(\iota) \rho_\iota(\mathbf{r}) d\iota , \qquad \int_{\Omega_\iota} \delta(\iota) d\iota = 1 , \tag{16c}$$

where $\delta(\iota) \equiv d\alpha/d\iota$ is the density of realisations, $\iota$ can be any proper parameter characterising $\alpha$, and the integrals above are taken over $\Omega_\iota$, the domain of $\iota$ variation of interest.

Note that in this way we introduce the new *concept, and the postulate, of the fundamental dynamic uncertainty* providing an explanation of origins of dynamical (in particular, quantum) chaos with the possibility of subsequent detailed description of the real chaotic system behaviour. The existence of such basic indeterminacy has been anticipated [1] as the necessary condition for the non-contradictory understanding of randomness in deterministic systems. It is important to emphasize, however, that in our approach this postulate is *naturally imposed* and specified by the discovered fundamental multivaluedness: eqs. (16) is the only reasonable issue permitting one to reconcile the existence of *many* equivalent realisations with the condition that one can observe *only one* of them. In fact, the axiom itself consists in the assumption that it is this modified form of the equation of motion (Schrödinger equation in our case) giving the multivaluedness that is the right and more general one as compared to the ordinary form which is relevant only to the special case of regular motion. To confirm this, one should verify the correspondence between the consequences of the assumption and the existing experimental, qualitative and quantitative, knowledge on chaos as well as its consistence with other fundamental principles of quantum mechanics.

To begin with, one may try to understand in more detail the physical meaning of eqs. (16). In the absence of perturbations other than $V_p$ the system would 'populate', with the probability from the set $\{\alpha_i\}$, one of the realisations (or, practically, groups of realisations) at the boundary or at the initial moment of time, and then it would remain within this realisation, theoretically, forever. In reality, however, as such additional perturbations (or noise) always exist, they will lead to 'spontaneous transitions' between different realisations. These transitions, however, are quite different from ordinary quantum-mechanical transitions or any other ones. To be consistent one should suppose that they occur, really or effectively, through a virtual intermediate state in which the system does not belong to any of the realisations (in fact, it virtually populates a realisation from another set determined by a new short-living system incorporating the complex system in question and the current configuration of the noisy perturbation). In other words, a noisy perturbation effectively 'knocks out' the system from the occupied realisation, and then it again 'chooses' a realisation at random, but with the probability from the same set $\{\alpha_i\}$. The last suggestion means that the noisy perturbation has no appreciable influence on the dynamical properties of the system determining its realisations and $\{\alpha_i\}$ or, in other words, that the mentioned virtual set of realisations is really virtual and 'perfectly mixing' with respect to the basic realisations. It seems to be a reasonable assumption: in the opposite case one should include at least a part of noisy perturbation into the main Hamiltonian (this may be considered as a definition of the boundary between the additional perturbation and the main Hamiltonian still leaving, probably, some freedom of choice).

Our main statement concerning the origin of quantum (and, in general, dynamical) chaos is that it is reduced to these transitions between different realisations of the effective Hamiltonian (or other relevant dynamical function) assisted by noise. It is important to emphasize that this role of noise is a significant, but not the crucial, one. The essential origin of chaos is the plurality of realisations creating, by definition, the effective instability, and it is a purely dynamical, non-stochastic effect (see also the discussion about the physical meaning of instability in section 5.III). Then the role of noise is to initiate (in the case of discrete realisations), or to amplify (for the continuous realisation distribution), a manifestation of this instability, the quasi-spontaneous transitions described above. It means, in particular, that the observed chaotic spectrum can be quite different from that of the applied noise and will depend on parameters, contrary to a non-chaotic system; it is determined by the dynamical properties of the system. On the other hand, it means that the role of noise can well be played by a quite regular additional perturbation incommensurate with the main one in its period or, in general, in its symmetry. It is thus an interesting possibility for another type of chaotic behaviour, characterised by a specific noise-free initiation of transitions between discrete realisations (it can be related to the amplifying influence of the zero-order potential nonlinearity, see the next section).

Generalising these observations, one may introduce the following symbolical representation (or definition) of the dynamics, $\mathcal{D}$, of an arbitrary system with whatever complex behaviour (one may hope that it is valid not only for periodically perturbed Hamiltonian systems, see the corresponding discussion in section 6.III):

$$\mathcal{D} = \{\{\mathfrak{R}_i\},\{\alpha_i\},N_\mathfrak{R}\} , \qquad (17)$$

where the quantities $\mathfrak{R}_i$, $\alpha_i$, and $N_\mathfrak{R}$ are defined above. Eq. (17) may be considered also as the statement (entering, in fact, the mentioned postulate of dynamic uncertainty) that the *set of realisations is complete* and therefore the dynamics is entirely determined by the right-hand side of eq. (17).

It is clear that there may exist two limiting manifestations of the phenomenon of dynamical chaos thus defined. In the first case, a noisy perturbation is sufficiently strong or, in other words, the separation between realisations is small enough. In this case we obtain typical highly irregular behaviour with pronounced explicit manifestations of randomness in the form of fast irregularities of motion, diffusion in the set of effective occupied states, etc. This regime is realised, in particular, in the semiclassical limit, when the typical separation between realisations is going to zero (see the end of section 2.2.I). This situation tends qualitatively to the classical picture of dynamical chaos. The characteristic features of the latter, such as instability, divergence of trajectories, and phase-space diffusion, are reduced, in quantum-mechanical terms, to frequent random transitions between closely separated realisations (where the nonlinearity of oscillations, if it exists, plays an important amplifying role, see the next section). The classical picture is therefore well reproduced. However, it is important to emphasize that the influence of additional perturbation (or noise) on a quasi-classical chaotic system is characterised by a small, but distinct, threshold which is just the quantum-mechanical feature explained by the finite realisation separation due, eventually, to the finite h value. It is already a manifestation of the quantum suppression of chaos in our sense (see the next paragraph) which can, in principle, be observed in some 'pure' (low-noise) semiclassical systems. Such threshold behaviour of semiclassical quantum chaos depending on the level of noise was indeed 'observed' in computer experiments [2,3], even though it is difficult to be sure that it can be attributed definitely and exclusively to the above mechanism (see the discussion of the relation to computer simulations in section 5.III).

The second limiting situation corresponds to a relatively large separation of realisations or, equivalently, to the relatively small noisy perturbation, when the 'spontaneous' transitions between realisations are rare, so that during the characteristic observation time they are hardly to occur. In this case we have a quasi-regular dynamics which experimentally may show practically the same features as the regular one including time reversibility, apparently total "quantum suppression of chaos", etc. We emphasize, however, that in reality it is a manifestation of the same quantum chaos mechanism, which should show up, for example, in the occurrence of more than one realisation for an *ensemble* of identical systems or experimental repetitions for the same system. In practice it may be difficult to separate the results of the repetitive measurements and the corresponding realisations from each other, which may manifest itself in the appearance of effective additional noises leading to the respective broadenings, etc. It is important that the limiting case considered can be realised in essentially quantum conditions, far from the semiclassical limit. This shows that dynamical chaos can well be compatible with quantum mechanics, even though in many essentially quantum situations it may manifests itself 'less irregularly', in the particular sense specified above.[*)] It is this specific mechanism of the *relative* diminishing of manifestations of the true dynamical randomness which gives *partial* quantum suppression of chaos in our approach. This mechanism has a very universal character: formally, it is always present when there is a chaotic regime in a quantum system (i. e. when $N_\Re > 1$). However, its 'usual domain of application' is centred around chaos in essentially quantum systems. In fact, this phenomenon replaces the *absolute* chaos suppression in quantum systems known from the existing theories of quantum chaos. Further on we shall meet with another, less universal, type of quantum suppression of chaos in time-independent Hamiltonian systems. It is interesting to note that the 'essentially quantum' limiting case considered can effectively take place also in the semiclassical situation, if only the level of noise is sufficiently low, and the experiments with excitation of atoms by electromagnetic field may be a relevant example [4]. In a situation of this type quantum suppression of chaos can be considerable even beyond its usual scope of applicability. This remark evidently returns us to the corresponding particular situation considered for the previous limiting case (see the end of the preceding paragraph).

---

[*)] This conclusion is supported and specified by consideration of the fractal structure of quantum chaos, section 4.IV.

# 3. Quantum chaos in Hamiltonian systems with periodic perturbation

To further specify the features of quantum chaos within the proposed formalism of FMDF (fundamental multivaluedness of dynamical functions), it is convenient to analyse different types of system behaviour depending on parameters. As follows from the main postulate, the global character of the system dynamics is determined by the number of distinct realisations for that system. To demonstrate the existing possibilities, we use the example of periodically perturbed system considered above and return to its graphical analysis, presented in Fig. 1.I and based on eqs. (13.I)-(15.I) (we start with the case of time-independent perturbation).

The realisations are determined by the branches of the function $V_{nn}(\varepsilon_{\sigma n})$ confined by a series of vertical asymptotes. As it follows from eq. (15a.I), the positions of the latter on the horizontal axis are determined by two characteristic energy intervals, $\varepsilon_1 = \Delta\varepsilon_\sigma$ and $\varepsilon_2 = 2\sqrt{\varepsilon_{\pi g \pi 0}(E - \varepsilon^*)}$, where $\varepsilon^* = \varepsilon_{\pi g \pi} \sin^2 \alpha_{g\pi} + \varepsilon_{g\pi n'}$. The generic chaotic behaviour, exemplified by Fig. 1(a).I, corresponds to $\varepsilon_2 \leq \varepsilon_1$. It is characterised by the maximum number of realisations $N_\Re = N_\Re^{\max}$. Moreover, the splitting of eigenvalues has a complex, irregular character: the 'normal' splitting due to the addition of degrees of freedom (the sum over $n'$ in eqs. (14.I)) is superimposed, in an entangled fashion, on the excessive, 'chaotic' one representing the fundamental multivaluedness.

If now the parameters change so that $\varepsilon_2$ increases, then at $\varepsilon_2 > \varepsilon_1$ one obtains another characteristic situation, illustrated by Fig. 1(b).I (for convenience and without any essential change, now we take into account three terms in the sum over $n'$ in eqs. (14.I), $N_{\pi'} = 3$). One can see that, whereas the total number of solutions is determined, formally, by the same rules as in the previous case (see eq. (9b.I)) and should remain unchanged for the same $N_\pi, N_{\pi'}$, they are now subdivided into the dense groups (realisations) of 'normal' solutions corresponding to ordinary dimensional splitting, these realisations being separated by larger intervals between them. The realisations here are similar to each other, and the higher is the ratio $\varepsilon_2/\varepsilon_1$, the more they are indistinguishable. Thus the two splittings, 'normal' and 'chaotical', are now well separated and not intermixed.[*] And what is most important, now it is difficult to distinguish formally different, but practically very similar, realisations one from another. All these arguments permit us to identify the point $\varepsilon_2 = \varepsilon_1$ as the (approximate) border between global chaos and regularity in the parameter space, the global chaos appearing at $\varepsilon_2 \leq \varepsilon_1$. This corresponds qualitatively to the predictions of classical description of chaos concerning the existence of the transition chaos-regularity in periodically perturbed systems [5-7]. In order to make a more detailed comparison, we express this condition of chaos onset through the parameters of a problem:

$$\Delta\varepsilon_\sigma \geq 2\sqrt{\varepsilon_{\pi g \pi 0}(E - \varepsilon^*)}$$

or (18)

$$E \leq (\Delta\varepsilon_\sigma)^2/(4\varepsilon_{\pi g \pi 0}) + \varepsilon^* \equiv E_c \approx (\Delta\varepsilon_\sigma)^2 d_\pi^2 m/(8\pi^2 h^2) ,$$

where $d_\pi$ is the perturbation period ($g_{\pi 0} = 2\pi/d_\pi$), and the last equality is valid if $\Delta\varepsilon_\sigma \gg \varepsilon_{\pi g \pi 0}$. In the semiclassical situation when $\Delta\varepsilon_\sigma \gg \varepsilon_{\pi g \pi 0}$ and $\Delta\varepsilon_\sigma = h\omega_\sigma$ ($\omega_\sigma$ is the classical oscillation frequency for the unperturbed system), one obtains

$$E_c = \omega_\sigma^2 d_\pi^2 m/(8\pi^2) . \tag{19}$$

---

[*] It means, by the way, that for this case one can impose a more exact limitation on the maximum number of realisations than the formal one, eq. (12b.I): $N_\Re^{\max} = N_\pi$; for the case of Fig. 1(b).I one have $N_\Re^{\max} = 4$ instead of $N_\Re^{\max} = 10$. Moreover, we have $N_\Re = N_\Re^{\max} = N_\pi$.

We see that this expression for the border 'global chaos' - 'global regularity' obtained by the ordinary semiclassical limit from our purely quantum-mechanical analysis contains only classical parameters. We shall call condition $E = E_c$ the *classical border of chaos*, even though in its general form, eq. (18), it can determine the onset of chaos in an essentially quantum situation. Now to compare it to the equivalent expression obtained within the classical analysis of the standard model (kicked oscillator), one should first pass to the conventional parameter $K$ [5-7] (for the details of the relation between the periodically perturbed system of the type considered here and the standard model see e. g. ref. [8], section 4.3). For the case of small harmonic oscillations in the unperturbed potential one easily obtains

$$K = m\omega_\sigma^2 d_\pi^2/(2E) \tag{20}$$

(see ref. [9], section 2.5, for details). Then the semiclassical limit of our condition for the global chaos onset, eqs. (18), (19), takes the form $K > K_c = 4\pi^2$. The classical estimations and computer calculations for the standard model and the related systems give $K_c \approx 1$ [5-7]. Thus our quantum-mechanical estimations, based on the formalism of FMDF, give an apparent discrepancy with the corresponding classical results only in numerical coefficient.

This difference can be easily understood and effectively eliminated if one takes into account the nonlinear character of the unperturbed oscillations and the conditional character of the border chaos-regularity. The former means that the linear oscillation frequency in the above expression for $K$ should be, in fact, replaced by certain effective frequency, $\omega_\sigma \to \omega_\sigma/l$, where the constant $l$ depends on the form of the particular potential well and is of the order of several units (it characterises the frequency spectrum width of the nonlinear oscillator, i. e. its anharmonicity). Then condition (20) turns into

$$K > K_c = (2\pi/l)^2 , \tag{21}$$

which coincides with the classical result at a reasonable value of $l = 2\pi$.

Note that this amplification of chaos due to $l > 1$ (the corresponding domain of global chaos becomes larger) can be considered also as a result of the specific influence of nonlinearity of the unperturbed oscillations manifested as the effective decrease of the realisation separation (as is clear, for example, from the graphical analysis, cf. Fig. 1(a).I). The value of $l$ gives us the idea about the magnitude of this amplifying influence of nonlinearity on chaos. In particular, for highly anharmonic unperturbed oscillations (for example, for a steep-sided exponential form of the zero-order potential), one may have effectively $l \to \infty$, which gives $K_c = 0$, and we obtain an extremely chaotic system that never leaves the state of global chaos (it could belong, for example, to the class of K-systems studied in the classical theory of dynamical chaos, see e. g. [5]). In terms of our general representation of chaos as random transitions between quasi-regular realisations, it means that the effective realisation separation for such systems goes to zero, and there is no more threshold for random transitions, so that they 'happen all the time', and the (quantum) system finds itself in a continuous purely chaotic 'wandering'. This particular behaviour can be interpreted also as the extreme system sensitivity to small noises (parameter fluctuations), where even infinitesimal noise intensity is sufficient to change the global system state, and the quantum chaos suppression, discussed in the previous section, is inefficient. The situation considered is too special, however, to provide a general enough origin of such high sensitivity to noises, a fundamental property of the classical chaotic behaviour. In a general quantum case, when the effective realisation separation is finite, we do have the partial suppression of chaos. Nonetheless, the latter can be somewhat attenuated by a universal enough mechanism involving the fractal structure of realisations and considered below, in section 4.III.

In connection to the condition (21) one may notice also that the transition points between different dynamical regimes thus determined can contain, by definition, some uncertainty expressed by a numerical coefficient. However, the latter should usually be rather close to unity and does not influence the physical relations obtained, even though the absolute quantitative accuracy is neither guaranteed. On the other hand, this uncertainty has its irreducible part related to the finite *width* of the transition chaos-regularity. The width of the considered classical border of chaos is determined by the variations of the quantities entering eqs. (18),(19) which depend much on the particular situation. If the relevant variation of the effective $\Delta \varepsilon_\sigma$ is expressed as $\Delta \varepsilon_\sigma / l'$ (or $\Delta \omega_\sigma = \omega_\sigma / l'$) with certain constant $l' \gg 1$, characterising the unperturbed potential anharmonicity, then the border width is $\Delta E_c \approx E_c / l'^2$. Therefore the more is the anharmonicity, the narrower is the transition to the developed chaos.

An important qualitative feature of this transition in essentially quantum systems is that it should have a pronounced step-wise character related to the sequential involvement of the individual energy levels in the chaotic dynamics. This is the *quantized chaoticity*: chaos appears by discrete 'portions' corresponding to particular energy levels. It should not be confused with a more general property of realisation discreteness: each realisation involves the totality of energy levels. It is clear that the unified notion of the classical border width, introduced above, can hardly be directly applied to such particular essentially quantum chaotic dynamics.

Note that the semiclassical condition, eq. (19), can be presented in another form revealing more physics. To obtain it, we introduce the effective time period of the perturbation, $T_\pi$. It emerges if we imagine that the total energy $E$ is entirely concentrated in the degrees of freedom $\mathbf{r}_\pi$, $E = mv_\pi^2/2$, and then $T_\pi$ is the time needed to cover one period, $d_\pi$: $T_\pi = d_\pi\sqrt{m/2E}$. Then eq. (19) is equivalent to a simple resonance relation, $T_\pi = T_\sigma$, or alternatively, $\omega_\pi = \omega_\sigma$, where $T_\sigma \equiv 2\pi/\omega_\sigma$ and $\omega_\pi \equiv 2\pi/T_\pi$. The global chaos will exist at $\omega_\pi < \omega_\sigma$ or $T_\pi > T_\sigma$, where, as was noticed above, $\omega_\sigma$ and $T_\sigma$ refer to certain effective values of the unperturbed potential parameters. In fact, the effective frequency $\omega_\sigma$ should represent the lower end of the conditional width of the frequency spectrum (i. e. it should correspond to sufficiently high bound states), in order that the above relation $\omega_\pi < \omega_\sigma$ extends automatically to the most part of the spectrum. It is clear that this would ensure the globality of chaos. This formulation of the condition for the global chaos onset seems to be equivalent to the ones known from classical mechanics [5,6], and similarly to them, it needs the support of computer simulations for its precise expression (providing e. g. the exact value of 1). It is interesting that this simple resonance interpretation of the classical border of chaos can be extended to the general quantum case, eq. (18). It is easily seen (cf. eqs. (3.I)) that the quantity $2\sqrt{\varepsilon_{\pi g \pi 0}(E - \varepsilon^*)}$ represents the main portion of the discrete energy transfer between the 'perturbation' degrees of freedom, $\mathbf{r}_\pi$, and those of the unperturbed problem, $\mathbf{r}_\sigma$ (see also section 5.III). Then if the effective (in fact, lower) energy-level separation, $\Delta\varepsilon_\sigma$, is at resonance with this energy transfer (at $E = E_c$), it means that the chaos criterion, eq. (18), is fulfilled for the majority of bound states and therefore it acquires the global character.

This interpretation provides a general explanation of the correspondence to classical results. The mentioned resonance conditions turn to zero the EP denominators (see eqs. (6.I), (14.I)) providing additional problem realisations. In the semiclassical situation these resonant denominators coincide (up to the factor h) with the analogous classical resonance conditions known to be directly involved in classical chaos [5-7]. This automatically ensures the equivalence of the ensuing conclusions including, for example, the resonance overlapping criterion [5-7]. Yet in the general case our expressions retain their purely quantum origin. This profound similarity of physical interpretation of chaos in our quantum-mechanical description to that in classical mechanics is a specific feature of the unreduced optical potential formalism, stemming from its basic properties (see section 5.III).

The resonance interpretation of chaos is related to another possible description of the transition to developed chaos. Namely, the global chaos onset is associated with *overlapping of the neighbouring realisations*, as is clear especially from Fig. 1(b).I and the accompanying explanations. Then the effective 1 value specifies the exact degree of this overlapping necessary for global chaos. This formulation does not seem to be the *direct* quantum-mechanical analogue of the classical criterion of resonance overlap but rather another, equivalent, interpretation of the condition (18).

It is important to mention another point of agreement between the results of our approach and those obtained within classical mechanics which is due, eventually, to the same intrinsic quantum-classical similarity within the EP formalism. It concerns the relation between chaos and regularity outside the classical border. It is clear from our description that within the domain of regularity chaos disappears asymptotically with the distance to the classical border. It means that some chaos always exists (the realisations are never exactly identical and the distance between them is always finite) and where it exists it is strong, but the relative number of such situations diminishes as one goes farther from the border. This is in good agreement with the well-known classical results [5,6,10]. The more detailed analysis of the expressions for EP, which will be described in the next paper (see also section 4.III), shows that this agreement can be extended to the conclusion that 'chaotically rich' features within the global regularity pass to the 'stochastic layer' and 'stochastic web' in the semiclassical limit, with the agreement between classically and quantum-mechanically obtained expressions for the width of the layer; as the transition to global chaos is approached, these features grow to cover finally all the accessible domain.

Consider now the second generic possibility concerning the transition chaos-regularity. To discover it note that while analysing above eq. (15a.I) determining the asymptote positions, we have neglected, in fact, the possibility that $E < \varepsilon^*$ under the root. Now if it is realised, the number of asymptotes is no more constant. When the parameter E diminishes from high values and becomes less than one of the discrete values of $\varepsilon^* = \varepsilon_{\pi g_\pi} \sin^2 \alpha_{g_\pi} + \varepsilon_{g_\pi n'}$, the corresponding pairs of asymptotes 'close' leading, in principle, to the disappearance of the respective branches of the function $V_{nn}(\varepsilon_{\sigma n})$ and thus of a realisation of a problem. Finally, if $E$ becomes so small that

$$E < E_q , \quad E_q = \overset{0}{\varepsilon}_{g_\pi 1} \cong \varepsilon_{\sigma 0} \equiv \min(\varepsilon_{\sigma n}) , \qquad (22)$$

where $\overset{0}{\varepsilon}_{g_\pi 1}$ is the first excited energy level in the set $\{\overset{0}{\varepsilon}_{g_\pi n}\}$, then the number of points of intersection of the two curves attains its minimum value corresponding to the ordinary dimensional splitting. In this case $N_\pi' = 1$, and, as follows from eq. (9b.I), $N_r^{\max} = N_\pi + 1 = N_\pi' + 1 = N_r^0.$[*] It means that we have only one realisation of a problem, and there is no any chaos at all. We shall call $E_q$ the *quantum border of chaos*, as opposed to the classical border of chaos, $E_c$, introduced above and coinciding, in the semiclassical limit, with the classically obtained quantity. Contrary to this, the existence of quantum border is a purely quantum-mechanical effect: formally $E_q \to 0$ when $h \to 0$; as follows from eq. (22), it is of the order of the lower energy level for the unperturbed potential, $\varepsilon_{\sigma 0}$. It represents another, more particular, but still rather general, case of the quantum suppression of chaos. The first case was described above and corresponds to partial chaos suppression due to the finite realisation separation. In contrast to this, the suppression of chaos under condition (22) is complete, and it is also a purely quantum-mechanical effect due to the finite value of the lower energy state.

---

[*] At these low energies even this minimal splitting can easily be suppressed, $N_r < N_r^0$, which corresponds to the existence of impenetrable barriers, forbidden energy zones, etc.

Note that the considered type of quantum suppression of chaos exists also at the other side of the quantum border, within the regime of global chaos, and it is *there* that it is only partial and diminishing with distance to the border. This is explained by the gradual appearance of asymptotes in our graphical representation and the corresponding new 'chaotic' realisations. Then one can deduce from eq. (15a.I) that this type of quantum suppression becomes much less pronounced if

$$E > E_q^*, \ E_q^* \cong \varepsilon_{g\pi 1}^0 \gg E_q . \tag{23}$$

The result obtained determines also the large-scale width of the quantum border which is $\Delta E_q = E_q^* - E_q \approx E_q^*$. Note that this transition should evidently have a step-wise form reflecting the discreteness of energy levels, which is a manifestation of the quantized chaoticity that can exist also near the classical border (see above in this section).

It is interesting to note, by the way, that it can explain the regularity of the elementary complex constituents of matter like nuclei and atoms. The existence of this property is not evident because each of the particles in such elementary agglomerate, e. g. an electron in atom, moves in a very complex effective potential of other particles which should typically give the pronounced chaotic behaviour (we have seen that it can well exist in quantum dynamics!). Moreover, even the asymptotic disappearance of chaos, like it happens beyond the classical border, would not help: unexcited matter seems to be absolutely regular (see, however, the next paragraph). The same concerns, of course, the partial quantum suppression of chaos described above. In contrast to this, condition (22) of the *complete* suppression of chaos agrees very well with this demand of absolute regularity, and the energy, for example, of atomic, or nuclear, ground state can self-consistently be just below the quantum border.[*]) Once being excited, an atomic electron leaves the domain of this total chaos suppression (it is true even apart from the chaotising effect of the exciting perturbation itself); this highlights a role of chaos in the processes of excitation of nuclei, atoms and solids which seems not to be recognised. From the other hand, this means that *any* state of a bound many-body system above the ground state is, in principle, chaotic.

---

[*]) Recall that our approach is applicable also for this case of chaos induced by symmetry breaking; physically, it is the asymmetrical part of the potential that plays the role of the effective 'periodic perturbation' with the period determined by the symmetric part.

This implication of the quantum border of chaos can be considered also as a basic rule governing the ground state formation in the bound many-particle quantum systems: if the current detailed configuration of the system which is just above its ground state does not correspond to the condition (22), it entails the presence of chaos giving rise to rearrangement of the constituents until the fulfilment of eq. (22). This induces, in particular, an interesting question about the existence (and even abundance?) of elementary bound quantum systems flickering in such never-ending search for a regular ground state which may not exist for them (a classical analogy for this specific quantum regime is provided by our solar system). This possibility is not subjected to the usual reduction to system relaxation to lower energy levels with the emission of quanta: the system in this particular regime is already in a specific ground state, it has already "tested" all the possibilities to go down. This does not preclude the existence also of a quasi-stable state of this kind relaxating eventually, after sufficiently long time, to the true ground state, chaotic or regular. It is easy to see that the possibility of these regimes is provided, in fact, by the very existence of the true quantum chaos. We leave the detailed investigation of such *chaotic ground and excited states* of the bound many-body quantum systems for special further studies.

Thus we have shown, using the postulate, and the quantum-mechanical formalism, of the fundamental multivaluedness, that more or less pronounced dynamical chaos can be generically observed in periodically perturbed Hamiltonian systems in the range of parameters

$$E_q < E < E_c \quad \text{or} \quad K_c < K < K_q , \tag{24}$$

where the borders $E_c$, $K_c$, and $E_q$ are determined respectively by eqs. (18), (21), and (22), while

$$K_q = (\Delta\varepsilon_\sigma)^2 d_\pi^2 m / (2\hbar^2 E_q) \tag{25}$$

(obtained by analogy to $K_c$). It is clear from eqs. (18), (22) that always $E_c > E_q$ and therefore periodically perturbed quantum system always possesses a regime of global chaos. For the considered case of time-independent perturbation the width of the interval of parameters corresponding to this regime is

$$\Delta E \equiv E_c - E_q = (\Delta\varepsilon_\sigma)^2 d_\pi^2 m / (8\pi^2\hbar^2) + \varepsilon^* - E_q \approx (\Delta\varepsilon_\sigma)^2 d_\pi^2 m / (8\pi^2\hbar^2) = (\Delta\varepsilon_\sigma)^2 / 4\varepsilon_{\pi g \pi 0} . \tag{26}$$

It can vary in a large scale from much smaller to much greater than such characteristic quantities as $\Delta\varepsilon_\sigma$, which may significantly influence the details of chaotic regime.

It is convenient to consider two possibilities determined by the relative magnitudes of the energy level separation, $\Delta\varepsilon_\sigma$, and another characteristic energy parameter, $\varepsilon_{\pi g \pi 0} = 2\pi^2\hbar^2 / m d_\pi^2$. If $\varepsilon_{\pi g \pi 0} \ll \Delta\varepsilon_\sigma$ then, as is seen from eqs. (18), (26), $\Delta E > \Delta \varepsilon_\sigma$, and the quantized chaoticity (the discreteness of the parameter $\varepsilon^*$) does not influence considerably the structure of the chaotic domain. This is the most frequent case, always occuring in the semiclassical regime, but it may happen also in essentially quantum situations. In the semiclassical limit it is characterised by a vanishingly small domain of the total quantum suppression of chaos at $E < \varepsilon_{\sigma 0}$, a rather large domain of the semiclassical global chaos at $E < E_c$, and an infinite interval of global regularity at $E > E_c$ with localised remnants of chaos asymptotically disappearing with growing $E$.

In the opposite case, $\varepsilon_{\pi g \pi 0} \geq \Delta \varepsilon_\sigma$, one has $\Delta E \ll \Delta \varepsilon_\sigma$, which leads, as is seen from eq. (15a.I), to multiplication of the domains of chaos analogous to the main one, eq. (24), but occurring intermittently with those of global regularity. Each of these domains of chaos begins at a discrete value of $\varepsilon^*$ ($= \varepsilon_{\mathbf{g}_\pi n'}$ for this case) and ends before the neighbouring higher value:

$$E_{qi} < E < E_{ci}, \quad i \geq 0, \quad E_{q0} \equiv E_q, \quad E_{c0} \equiv E_c, \qquad (27)$$

where $E_{qi}$ takes the consecutive values of $\varepsilon_{\mathbf{g}_\pi n'}$, and $E_{ci} = E_{qi} + \Delta E$. This regime represents a specific manifestation of chaos in essentially quantum systems. From one hand, chaos is considerably suppressed here because the domains of the pronounced chaos, eq. (27), are relatively narrow, $\Delta E \ll E_{qi+1} - E_{qi}$, and moreover, in each of them chaos is not complete: the number of realisations determined by the existing asymptotes (cf. Fig. 1.I) is generally far from its maximum. On the other hand, some degree of chaoticity is always present, especially for $E$ around excited states, with the exception of the interval $E < E_{q0}$. These details specify the above conclusion that excited states of quantum system are always chaotic; now we see the particular structure of chaos in an essentially quantum system, concentrated indeed around its excited states. One may regard this structure as a prototype of *chaotic broadening* of the excited states also for chaotic systems where the total energy is not fixed, and should be determined (the general eigenvalue problem). We see, from the above results, that the more pronounced is the quantum character of a system, the smaller are the intervals of chaos, - another manifestation of the partial quantum suppression of chaos. The regimes of this type can practically occur in systems with relatively deep and well separated bound states, heavy ion cores and nuclei providing the relevant examples.

It also reminds us of different regimes of partial chaos known from calculations in classical mechanics; we see now that the formalism of FMDF permits one to discover and study analogous diversity in quantum chaos which a priori is not smaller than that of classical chaos. Some particular regime and the corresponding character of quantum chaos may have its direct semiclassical analogue, like it was in the limit $\varepsilon_{\pi g \pi 0} \ll \Delta \varepsilon_\sigma$ above, but it is not always so as is demonstrated by the opposite case, $\varepsilon_{\pi g \pi 0} \geq \Delta \varepsilon_\sigma$. If now, starting from this latter situation, we increase the ratio $\Delta \varepsilon_\sigma / \varepsilon_{\pi g \pi 0}$ ($= \omega_\sigma m d_\pi^2 / 2\pi^2 \hbar$ by a semiclassical estimate), then all the intervals (27), the effective widths of the excited energy levels, grow with respect to energy-level separation, $\Delta \varepsilon_\sigma$, until they merge, and we return to the first case of the one large domain of developed chaos, eq. (24). This shows a mechanism of gradual change of the degree of quantum suppression of chaos depending on parameters.

Now we briefly summarise the corresponding results for the parameter dependence of chaos in the case of time-dependent perturbation, obtained in the same manner from eqs. (13.I), (14b.I), (15b.I) and the accompanying graphical analysis. In this case we have only classical border of chaos, i. e. global quantum chaos exists at

$$0 < \omega_\pi < \omega_{0c} , \qquad (28)$$

where

$$\omega_{0c} = \Delta \varepsilon_\sigma / \mathrm{h} = \omega_\sigma, \qquad (29)$$

$\omega_\sigma$ stands for the 'effective' classical oscillation frequency for the unperturbed potential, and the last equality applies to the semiclassical limit. Note that the 'resonance' interpretation of the global chaos onset, described above for the time-independent perturbation, becomes especially evident in the present case. Condition (28) can also be expressed through the 'parameter of chaoticity' $K$, in which form it coincides with eq. (21), where now

$$K = (2\pi/\mathrm{l})^2 (\omega_\sigma/\omega_\pi)^2 . \qquad (30)$$

We see that this case of time-dependent perturbation is somewhat simpler than the time-independent one. In particular, in the time-dependent case there is no second, quantum border of chaos (or, formally, it is infinitesimal, $\omega_{\pi\mathrm{q}} = 0$), neither an intermittent chaos regime as the one described above. The existence of these differences is quite natural and is due to the appearance of additional parameters and their relations in the time-independent formalism with respect to the other case: instead of the "one-dimensional" couple $t$-$\omega_\pi$ for the latter, one has the relation $\mathbf{r}_\pi$-$\{E,\mathbf{g}_\pi\}$ for the former, which provides more possibilities concerning, in particular, parameters and their relations. At the same time it is important to note that, apart from these particular distinctions, the two cases provide the same physical results in their most important parts: the existence of the true dynamical chaos in quantum system with periodic perturbation which passes to its classical analogue in the limit $\mathrm{h} \to 0$, including the border chaos-regularity. This similarity of the main physical results for the two cases has already been known from the classical chaos studies (see e. g. [8]), and now it finds thus its quantum-mechanical extension.

The diversity of quantum chaos regimes can be further illustrated by two more particular, but still important, patterns obtained within the same formalism. The first one concerns the quasi-free motion of a system above the periodic potential barriers. To avoid unnecessary complications, consider the case of effectively one-dimensional periodic unperturbed potential, $V_0(\mathbf{r}_\sigma) \equiv V_0(x)$. While analysing this type of chaos, one can follow the same scheme as the above one for the bound states, because in quantum-mechanical terms the above-barrier motion is also characterised by discrete energy zones. Then we arrive at the conclusion that there should exist, for this case also, the classical border of chaos determined by the same expressions, eqs. (18)-(21), (29), where one should substitute for $\Delta \varepsilon_\sigma$ the distance between the neighbouring energy zone extrema. An estimate for the latter can be obtained in the limit of weak periodic potential influence ("quasi-free electrons" in the solid state theory, see e. g. [11]) approximately valid for higher above-barrier states and/or relatively narrow and small barriers. In this model one has quasi-parabolic zones, $\varepsilon_\sigma(k_\sigma) \approx \mathrm{h}^2 k_\sigma^2/2m$, and, respectively, $\Delta \varepsilon_\sigma = 2\mathrm{h}^2 k_\sigma \Delta k_\sigma/m$, where $\Delta k_\sigma = \pi/d_\sigma$ is half the symmetric Brillouin zone ($d_\sigma$ is the period of $V_0(x)$). Combining it with eq. (18), one arrives at the following expression for the boundary between global chaos and global regularity (time-independent perturbation):

$$E_c = \left(\frac{d_\pi}{d_\sigma}\right)^2 \varepsilon_\sigma . \tag{31}$$

We see, once more (cf. eq. (19)), that the result, obtained quantum-mechanically, contains only classical quantities ($\varepsilon_\sigma$ is of the order of the amplitude of $V_0(x)$). This is not surprising in itself because the approximation used to obtain it corresponds well to the semiclassical limit. What is more interesting is that this result can be given a consistent interpretation in purely classical terms. Indeed, it is easy to see that condition (31) corresponds to the resonance between the periodic motion in zero-order potential $V_0(x)$ and the perturbation. This means that the time needed to traverse one period of $V_0(x)$ during the quasi-free motion of a system, $T_\sigma$, is equal to the effective time period of perturbation, $T_\pi = d_\pi \sqrt{m/2E}$. In the semiclassical situation the same result can be obtained in a slightly different way applicable to arbitrary potential $V_0(x)$. To reveal it, it is sufficient to note that in this case $\Delta \varepsilon_\sigma = \mathrm{h} \omega_\sigma' = 2\pi \mathrm{h}/T_\sigma'$, which transforms the quantum-mechanical condition of the global chaos, eq. (18), into the relation $T_\sigma' < T_\pi$, or $\omega_\sigma' > \omega_\pi = 2\pi/T_\pi$. We ascertain thus that in this form it coincides with the above 'resonance' interpretation of the global chaos onset for the bound states. It demonstrates the invariance of the proposed quantum-mechanical chaos description with respect to the motion type (bound or free).

The second particular type of chaotic Hamiltonian dynamics is also associated with the above-barrier motion, but it has purely quantum character without classical analogues. To discover it recall that chaotic behaviour is related to resonance EP denominators, eqs. (14.I). Considering those resonances, we have taken close values of $\varepsilon_{gn}^0$ and $\varepsilon_{\sigma n}$ separated typically by $\Delta \varepsilon_\sigma$, which corresponds to the largest matrix elements in the EP numerator and gives, in the semiclassical limit, classical resonance and classical chaos. However, another possibility can exist for these matrix elements to be relatively large, but now with a large difference between the zero-order energy levels ($\varepsilon_{gn}^0$) and those for the full problem ($\varepsilon_{\sigma n}$). Indeed, the matrix elements are large when the two differences, $[\varepsilon_{gn}^0 - V_{\mathbf{g}\pi}(\mathbf{r}_\sigma)]$ and $[\varepsilon_{\sigma n} - V_{\mathrm{eff}}(\mathbf{r}_\sigma)]$, taken for $\mathbf{r}_\sigma$ within the relatively flat extrema of the potentials, are close between them, with the deviations of the order of $\Delta \varepsilon_\sigma$. This means that $(\varepsilon_{gn}^0 - \varepsilon_{\sigma n})$ and $[V_{\mathbf{g}_\pi}(\mathbf{r}_\sigma) - V_{\text{eff}}(\mathbf{r}_\sigma)]$ should be close up to $\Delta \varepsilon_\sigma$. Such situation can be achieved in two physically different ways: first, if $(\varepsilon_{gn}^0 - \varepsilon_{\sigma n})$ and $[V_{\mathbf{g}_\pi}(\mathbf{r}_\sigma) - V_{\text{eff}}(\mathbf{r}_\sigma)]$ are both of the order of $\Delta \varepsilon_\sigma$, and this leads to the 'classical' chaos; second, if these two differences are allowed to take arbitrary values, but close between them up to $\Delta \varepsilon_\sigma$, and this provides a particular, essentially quantum, type of chaos. The fact that this second possibility can be realised, in principle, as easy as the first one, is confirmed by eqs. (14.I) from which it follows that in the second case the EP amplitude differs from the one for the unperturbed potential approximately by $2\sqrt{\varepsilon_{\pi g_\pi} E} = 2\pi \hbar g \sqrt{2E/m}/d_\pi = \hbar \omega_\pi g$, where $g$ is a (small) non-zero integer, and $\omega_\pi = 2\pi\sqrt{2E/m}/d_\pi$ is the effective time period of the perturbation, introduced above. It is easily seen that the total number of these particular EP configurations equals to $N_\pi + 1 = N_\Re^{\max}$. One of them represents the 'main' group of realisations which passes smoothly to the unperturbed potential when chaos disappears; typically the EP parameters in this group are of the same order of magnitude as those for $V_0(\mathbf{r}_\sigma)$. There are also $N_\pi/2$ pairs of configurations with the respective modifications of the EP amplitude by $\pm \hbar \omega_\pi g$; these realisations can provide deviations from the unperturbed characteristics by arbitrary large magnitudes depending on parameter values. Indeed, any real situation corresponds to a fixed 'effective' value of $\hbar$, while $\omega_\pi$ can be made arbitrary large.

One can formulate thus two physically important conclusions concerning this type of chaos. First, it is reduced to transitions between the realisations with the essentially different EP parameters, even in the semiclassical regime; the characteristic EP amplitude can suddenly change, during such transition, by a quantity which can take, in principle, whatever high (positive or negative) values depending on parameters. In particular, a potential barrier can be transformed into effective well, and vice versa. Second, these EP amplitude variations, as well as the corresponding chaotic behaviour in general, represent an essentially quantum effect, without any classical analogue. We see that, surprisingly, quantum mechanics can provide not only a suppression, but also a specific quantum source, of chaos! This source may play an important complementary role with respect to 'ordinary' chaotic regimes considered above. Indeed, we have seen that chaos becomes fully developed within the classical border of chaos, $\omega_\pi < \omega_\sigma$, disappearing asymptotically at $\omega_\pi \gg \omega_\sigma$, while the considered regime of 'quantum jumps' is activised just at $\omega_\pi > \omega_\sigma$. Moreover, there can evidently be no quantum suppression for this type of chaos, and the corresponding dramatic EP changes, becoming especially pronounced just at $\omega_\pi \gg \omega_\sigma$, can be used for the unambiguous detection of quantum chaos (specification for a particular physical system can be found in [9]). This possibility depends also on the 'density of realisations', see eqs. (16), which can be relatively small for quantum jumps, especially in the semiclassical regime. Note that the estimations and conclusions obtained can be verified and specified [9] with the help of a simple model mentioned at the end of section 2.2.I.

One of the particular consequences of this regime of 'essentially quantum' chaos with its large EP variations concerns the evident influence on the process of quantum-mechanical tunnelling. Namely, the latter can be considerably facilitated when, due to quantum jumps, the unperturbed potential barrier effectively diminishes or even changes to a well. Of course, this phenomenon could be anticipated to occur already within the 'ordinary' quantum chaos regimes due to relatively small, but permanent, EP variations between different realisations. However, in that case it can only manifest itself as a multi-stage diffusion-like process. Contrary to this, tunnelling induced by quantum jumps can happen 'instantaneously' and may be inhibited only because of their lower probability. In any case, these two forms of quantum chaos implication in tunnelling, that can be called *chaotic tunneling*, should typically occur in the different parameter domains separated by the classical border of chaos, $\omega_\sigma = \omega_\pi$, thus facilitating their experimental separation. Yet another mechanism of chaotic tunnelling essentially involving the fractal structure of quantum chaos is considered below, in section 4.III.

In the case of *'chaotic jump tunnelling'* one can propose a simple estimate of the magnitude of the effect: according to the general paradigm, eqs. (16), it is characterised simply by the total probability of the corresponding realisations. Specifically, chaotic jump tunnelling at the bound-state energy level $\varepsilon_\sigma$ for a particular barrier $\beta$ is characterised by the barrier transmission probability $p_\beta(\varepsilon_\sigma)$:

$$p_\beta(\varepsilon_\sigma) = \sum_{i \in A_\beta(\varepsilon_\sigma)} \alpha_i ,$$

where $A_\beta(\varepsilon_\sigma)$ is the set of realisations for which the $\beta$-th EP barrier height is less than $\varepsilon_\sigma$. The overall significance of the jump tunnelling through the barrier $\beta$ can be characterised by the total negative-jump probability $P_\beta$:

$$P_\beta = \sum_{i \in A_\beta^-} \alpha_i ,$$

where $A_\beta^-$ is the set of realisations for which the $\beta$-th EP barrier height is less than that for the unperturbed potential by more than $\approx \Delta\varepsilon_\sigma$.

The notion of "chaos-assisted tunnelling" has been also introduced recently [12] within the mixed classical/quantum analysis coupled to computer simulations and the random-matrix hypothesis, though without any hint on a 'true' quantum chaos. In our approach, on the contrary, the intrinsic randomness forms the necessary basis for this specific kind of tunnelling. We may suppose, however, that in both cases one deals with physically similar phenomena though analysed within rather different descriptions. This difference hampers the direct definitive comparison of the respective findings which could be realised with further development of both approaches.

We finish this section by noting that our method for quantum chaos analysis and the results obtained can be directly applied to, or easily specified for, the description of particular chaotic quantum Hamiltonian systems with periodic perturbation occurring in different fields of physics. This possibility is related, in part, to the fact that the generalised kicked oscillator considered above is reduced, for the particular potential and perturbation, to the well-known kicked rotor, or standard, model [7] which, in its turn, was shown to be relevant to many real physical systems (see also [5,6]). The general case analysed within our approach leads to more realistic approximations while describing the chaotic behaviour of these particular systems. The scheme presented above provides the complete description of a quantum chaotic system up to the measurable quantities (see e. g. eqs. (5.I)-(8.I), (16)), and we have demonstrated above that many significant features of the chaotic behaviour can be deduced from it already by analytical methods. Of course, these results can be verified and specified by subsequent application of numerical calculations which is out of the scope of the present paper centred around the fundamental problem of origin of quantum chaos.

One particular application of the proposed formalism concerns the manifestations of dynamical (quantum) chaos in charged particle channeling in crystals, the case for which our concept of quantum chaos was first formulated [9]. The general description, presented here, can be directly used to study this system by considering $r_\sigma$ to be the vector of the transverse coordinates of the channeling particle (with respect to the channel direction) and $r_\pi$ to be the vector of its longitudinal coordinates. By doing so one arrives to a number of experimentally verifiable predictions expressed in terms of this particular problem [9]. An interesting special case involves chaotical dynamics of analysing electrons used in the high-voltage electron microscopy of solids. The predicted chaos provides a basically unavoidable source of image unsharpness in this method of analysis, and it is just the case of the true chaos in an essentially quantum regime (see [9] for more details). Another evident group of applications concerns the role of chaotic tunnelling in the processes of particle reflection from surfaces or the high-energy particle beam deflection by deformed crystal. In a similar fashion, particular results can be obtained for many other different cases of chaotic behaviour in wave, or particle, scattering in periodic structures.